\documentclass[12pt]{iopart}

\usepackage{iopams} 
\usepackage[breaklinks=true,colorlinks=true,linkcolor=blue,urlcolor=blue,citecolor=blue]{hyperref}
\usepackage{rotating}
\usepackage{graphicx}
\usepackage{epsfig}
\usepackage{epstopdf}

\begin{document}

\pagenumbering{arabic}

\title{Study of the second magnetization peak and the pinning behaviour in Ba(Fe$_{0.935}$Co$_{0.065}$)$_2$As$_2$ pnictide superconductor}
\author{Shyam Sundar$^1$, J. Mosqueira$^2$, A. D. Alvarenga$^3$, D. S\'o\~nora$^2$, A. S. Sefat$^4$ and S. Salem-Sugui Jr.$^1$}

\address{$^1$Instituto de Fisica, Universidade Federal do Rio de Janeiro, 21941-972 Rio de Janeiro, RJ, Brazil}
\address{$^2$LBTS, Dept. Fisica de Particulas, Universidade de Santiago de Compostela, E-15782, Spain}
\address{$^3$Instituto Nacional de Metrologia Normalizac\~ao e Qualidade Industrial, 25250-020 Duque de Caxias, RJ, Brazil}
\address{$^4$Oak Ridge National Laboratory, Oak Ridge, TN 37831, USA}

\vspace{10pt}
\ead{shyam.phy@gmail.com}

\vspace{10pt}
\begin{indented}
\item[] August 2017
\end{indented}

\begin{abstract}
Isothermal magnetic field dependence of magnetization and the magnetic relaxation measurements were performed for $H$$\parallel$c axis on single crystal of Ba(Fe$_{0.935}$Co$_{0.065}$)$_2$As$_2$ pnictide superconductor having $T_c$ = 21.7 K. The second magnetization peak (SMP) for each isothermal $M(H)$ was observed in a wide temperature range from $T_c$ to the lowest temperature of measurement (2 K). Magnetic field dependence of relaxation rate $R(H)$, shows a peak (H$_{spt}$) between H$_{on}$ (onset of SMP in $M(H)$) and H$_p$ (peak field of SMP in $M(H)$), which is likely to be related with a vortex-lattice structural phase transition, as suggested in literature for similar sample. In addition, the magnetic relaxation measured for magnetic fields near H$_{spt}$ show some noise which might be the signature of the structural phase transition of the vortex lattice. Analysis of the magnetic relaxation data using Maley's criterion and the collective pinning theory  suggests that the second magnetization peak (SMP) in the sample is due to the collective (elastic) to plastic creep crossover, which is also accompanied with a rhombic to square vortex lattice phase transition. Analysis of the pinning force density suggests single dominating pinning mechanism in the sample and is not showing the usual $\delta$l and $\delta T_c$ nature of pinning. The critical current density ($J_c$) estimated using the Bean's critical state model is found to be 5 $\times$ 10$^5$ A/cm$^2$ at 2 K in the zero magnetic field limit. Surprisingly, the maximum in the pinning force density is not responsible for the maximum value of the critical current density in the sample.
\end{abstract}

\vspace{2pc}


\maketitle                        

\section{Introduction}

Superconductivity in iron-based superconductors (IBS) was discovered in 2008 and soon after it has been realized that the vortex dynamics in this class of superconductors is of great importance due to their moderately high superconducting transition temperature ($T_c$) \cite{an08}, high upper critical field ($H_{c2}$) \cite{sen08, jar08}, smaller anisotropy than cuprates \cite{yua09, alt08}, high intergrain connectivity \cite{tak11b, wei12} and low Ginzburg number ($G_i$) \cite{ele17}. These salient features of IBS make them suitable for technological applications \cite{asw10, wei16, pur15, tos15, mis16} and hence intensive research is still going on to study the vortex dynamics in these systems \cite{ele17}. In  past years, a few pnictide superconductors have been recognized as potential materials for high field applications such as Co and K-doped BaFe$_2$As$_2$ (Ba-122) pnictide superconductors \cite{wei12, kaz11, tak11a, tog12}. Beside the technological interest, the vortex dynamics in the pnictide superconductors is also very interesting because of the different phases that exist in the vortex phase diagram. Among these different phases, the second magnetization peak (SMP), also known as the fish-tail effect, has been widely studied in different systems and is still a topic of intense research \cite{wei16, shy17}. The SMP in superconductors is observed in the isothermal $M(H)$ below $T_c$, and is associated with a peak in the magnetic field ($H$) dependence of the critical current density ($J_c$). In the literature, the SMP behaviour in different superconductors is explained in terms of an order-disorder transition \cite{miu12, joh14}, elastic to plastic creep crossover \cite{wei16, said10}, vortex lattice phase transition \cite{kop10, pra11} and is still under investigation for a few compounds \cite{said13, said11}. 

The superconductivity in Co-doped Ba-122 was observed by Sefat et. al \cite{sef08} just after the discovery of iron based superconductors \cite{kam08}. Later, Prozorov et. al. \cite{pro08} constructed the vortex-phase diagram and observed the SMP behaviour in Ba(Fe$_{0.93}$Co$_{0.07}$)$_2$As$_2$, which is described in terms of crossover from collective to plastic creep \cite{pro08, bin10}. Further, Kopeliansky et. al. \cite{kop10} suggested that the SMP in Ba(Fe$_{0.925}$Co$_{0.075}$)$_2$As$_2$ is associated with a vortex lattice structural phase transition. However, they also argued that a crossover in the pinning mechanism (collective to plastic) may also be accompanied with the vortex lattice structural phase transition as it has also been observed in LaSCO superconductor \cite{ros05}. In the literature, a variety of superconductors such as LiFeAs \cite{pra11}, ErNi$_2$B$_2$C \cite{esk97}, YNi$_2$B$_2$C, LuNi$_2$B$_2$C \cite{esk97-2}, YBa$_2$Cu$_3$O$_7$ \cite{bro04}, CeCoIn$_5$ \cite{esk03} and Sr$_2$RuO$_4$ \cite{ray14} show a structural phase transition in the vortex lattice. These studies motivated us to choose a Ba(Fe, Co)$_2$As$_2$ sample with a Co-content similar to those used by Prozorov et. al. \cite{pro08} and Kopeliansky et. al. \cite{kop10} to investigate whether a clear signature of pinning crossover (collective to plastic) exists within the vortex lattice structural phase transition picture. Hence, we performed a detailed study of isothermal $M(H)$ and magnetic relaxations $M(t)$ for $H$$\parallel$c axis in Ba(Fe$_{0.935}$Co$_{0.065}$)$_2$As$_2$ pnictide superconductor (similar composition as used by Prozorov et. al. \cite{pro08} and Kopeliansky et. al. \cite{kop10}) below $T_c$ and analyzed the data using Maley's analysis \cite{mal90} and collective pinning theory \cite{fei89}. Our results clearly suggest that the SMP in this compound is associated with the crossover in the creep mechanism (collective to plastic) which is likely to occur above a rhombic to tetragonal vortex lattice structural phase transition in the sample. We also estimated the $J_c$ using Bean's critical state model \cite{bean64} and compared it with other studies. The pinning force density is also analyzed using models developed by Dew-Hughes \cite{dew74} and Griessen et. al. \cite{gri94}.

\section{Experimental Details}

In the present work, we performed a detailed study on Ba(Fe$_{0.935}$Co$_{0.065}$)$_2$As$_2$ single crystal to understand the nature of SMP and pinning behaviour in the sample. The single crystal was synthesized using a flux method described in previous reports \cite{sef08, sef09, sef13}. The chemical doping of the single crystal was checked by energy-dispersive X-ray spectroscopy, and confirmed through the refinement of lattice parameters with X-ray diffraction. The sample used in the study is a 3.168 mg platelet with a surface of 6.84 mm$^2$. These parameters (together with the density resulting from the lattice parameter \cite{har11}) were used to estimate the average thickness of the sample as $\sim$ 62~$\mu$m. Through microscopic inspection, it is found that the crystal surface is somewhat irregular and uniformity of the thickness may be estimated within $\sim$ 10\% accuracy. The measurements were performed with a Quantum Design's squid-based magnetometer (model MPMS-XL). The sample was installed in a quartz sample holder with the Fe-As layers perpendicular to the applied magnetic field. The hysteresis cycles were performed after zero-field cooling (ZFC) the sample to the target temperature, and then charging the magnetic field with the so-called hysteresis (non persistent) mode. The magnetic relaxation measurements were also performed after ZFC to the desired temperature, and then charging the successive magnetic fields with the hysteresis mode. For each magnetic field the data were collected during $\sim$ 80 min while the field was held stable in the non-persistent mode.

\section{Results and Discussion}

\begin{figure}[h]
\centering
\includegraphics[height=8cm]{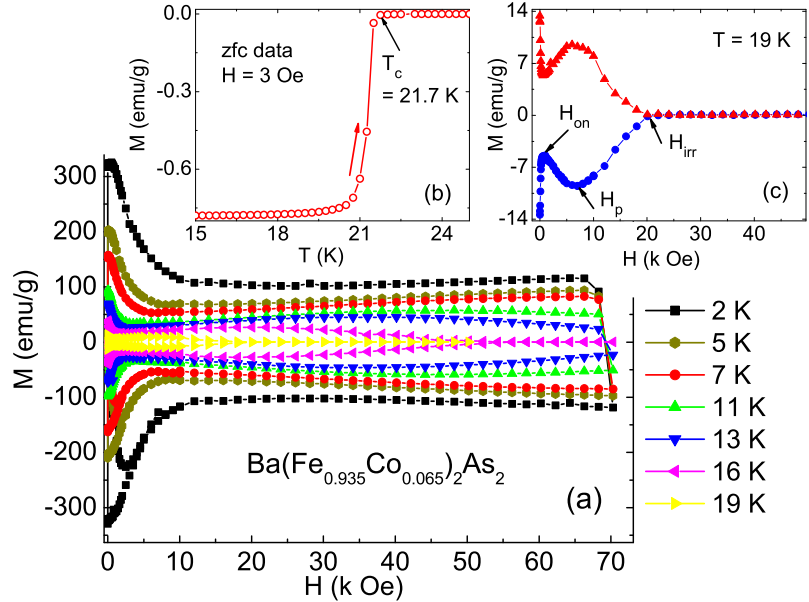}
\caption{\label{fig:MH_MT} (a) Isothermal magnetic field dependence of the magnetization, $M (H)$, at selected temperatures well below $T_c$. (b) Temperature dependence of zfc magnetization, $M(T)$, at $H$ = 3 Oe, showing the onset of superconducting transition at 21.7 K. (c) The characteristic fields $H_{on}$, $H_p$ and $H_{irr}$ are well defined for $M(H)$ measured at T = 19 K.}
\end{figure}

Figure \ref{fig:MH_MT}a shows the isothermal magnetic field dependence of the magnetization, $M(H)$, for selected temperatures well below $T_c$. The measured isothermal $M(H)$ show the well defined second magnetization peak (SMP) at all temperatures from near $T_c$ to the lowest temperature of our measurement (2K). However, in few superconductors, the SMP exists only in a limited temperature range below $T_c$ \cite{shy17, tam93, yes94}. Fig. \ref{fig:MH_MT}b shows  the temperature dependence of zfc magnetization, $M(T)$, measured with a 3 Oe magnetic field showing the onset of superconducting transition at 21.7 K. The characteristic fields $H_{on}$, $H_p$ and $H_{irr}$ are well defined in Fig.\ref{fig:MH_MT}c. Magnetic relaxation measurement is a widely used technique to examine the characteristics of SMP \cite{wei16, shy17, said10, said13}. Similarly, we performed magnetic relaxation, $M(t)$, measurements on selected isothermal $M(H)$, for about 80 minutes to study the SMP in present sample. The observed magnetic relaxation shows the usual logarithmic time dependence, ln$\mid$$M$$\mid$ $\sim$ ln $t$ and the relaxation rate ($R$ = dln$\mid$$M$$\mid$/dln$t$), was obtained using the plot of ln$\mid$$M$$\mid$ vs. ln$t$.

\begin{figure}[h]
\centering
\includegraphics[height=8cm]{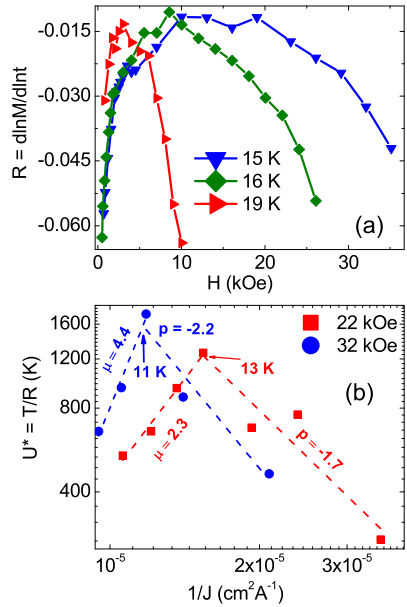}
\caption{\label{fig:RH_U} (a) Magnetic field dependence of relaxation rate, $R = dlnM/dlnt$, at constant temperatures. Each curve shows a peak behaviour in $R(H)$. (b) Inverse current density ($1/J$) dependence of activation energy ($U^*$) measured for different isothermals.}
\end{figure}

Figure \ref{fig:RH_U}a shows the $H$ dependence of relaxation rate ($R$) at different temperatures. In each isothermal $R(H)$ curve, the absolute value of $R$ decreases initially then increases for higher fields and shows a peak at intermediate fields. This peak in each isothermal $R(H)$ suggests some correspondence with the respective $M(H)$ and is lying in the peak effect region ($H_{on} < H < H_p$), which is discussed further. we have also obtained the temperature dependence of the relaxation rate and plot the corresponding activation energy, $U^* = T/R$, as a function of the inverse current density $1/J$ ($J$ is obtained using Bean's model). These methods has already been exploited previously to examine the vortex dynamics in different pnictide superconductors \cite{wei16, shy17, said10, tos12}. The relation between $J$ and the activation energy $U^*$ is defined as $U^* = U_0(J_c/J)^{\mu}$, where $\mu$ and $J_c$ (critical current density) depend on the dimensionality and size of the vortex bundles under consideration \cite{fei89}. The exponent $\mu$ may be extracted using the double logarithmic plot of $U^*$ vs. $1/J$, which is shown in Fig. \ref{fig:RH_U}b. In the present case, the $\mu$ values are found to be about 2.3 and 4.4 for 22 kOe and 32 kOe respectively. However, for 3D systems, $\mu$ values were predicted to be 1/7, 3/2, 7/9 for single-vortex, small-bundle, and large-bundle regimes, respectively \cite{fei89, gri97}. Further, the exponent for higher temperatures side is estimated to be about -2, which is also different from the predicted value ($p$ = -1/2) for plastic creep mechanism \cite{abu96}. In-spite of the difference between the obtained and the predicted exponents, the $U^*$ vs. $1/J$ curves for both fields suggest a crossover in the pinning mechanism. However, in our recent study \cite{shy17}, it has been suggested that the above method must be used with caution to study the SMP behaviour in superconductors. 

\begin{figure}[h]
\centering
\includegraphics[height=8cm]{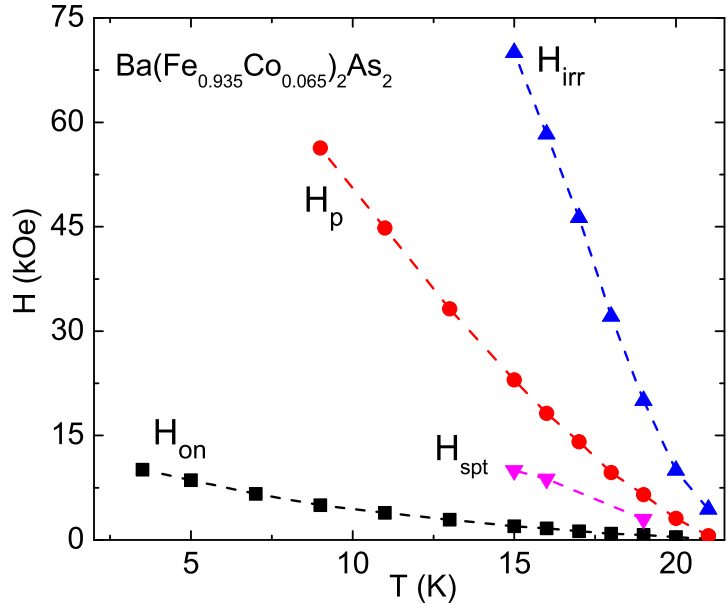}
\caption{\label{fig:HT-diagram} $H$-$T$ phase diagram for the sample used in the present study.}
\end{figure}

In Fig. \ref{fig:HT-diagram}, the $H$-$T$ phase diagram shows the different characteristic fields, $H_{on}$, $H_p$, $H_{irr}$ and $H_{spt}$, associated with the SMP, the irreversibility line, and the peak in $R$ vs. $H$. It is shown that both $H_{on}$ and $H_p$ exist well below the $H_{irr}$ line in the whole temperature range below $T_c$. It should be noted that the $H_{spt}$ line falls between $H_{on}$ and $H_p$ as has been previously observed by Kopeliansky et al. \cite{kop10} for Ba(Fe$_{0.925}$Co$_{0.075}$)$_2$As$_2$. Similar to the La$_{2-x}$Sr$_x$CuO$_4$ system \cite{ros05}, Kopeliansky et al. \cite{kop10} explained the observed $H_{spt}$ line in terms of a vortex-lattice structural phase transition, from a rhombic to square lattice, caused by the softening of the vortex lattice. Recently, similar vortex lattice structural phase transition (rhombic to square) was also suggested in  BaFe2(As$_{0.68}$P$_{0.32}$)$_2$ superconductor by Salem-Sugui et. al. \cite{said15}. It suggests that the $H_{spt}$ line in the present case might also be associated with a similar vortex-lattice structural phase transition and such phase transition may be accompanied with a crossover from elastic to plastic creep. However, no direct evidence has been observed to support the claim made by Kopeliansky et al. \cite{kop10} and Salem-Sugui et. al. \cite{said15}. In addition, vortex imaging studies revealed highly disordered (no long range order) vortex-lattice configuration in different Co-doped BaFe$_2$As$_2$ \cite{esk09, yin09} and other pnictide superconductors \cite{vin09}, which makes it difficult to observe a clear signature of structural phase transition.       

\begin{figure}[h]
\centering
\includegraphics[height=8cm]{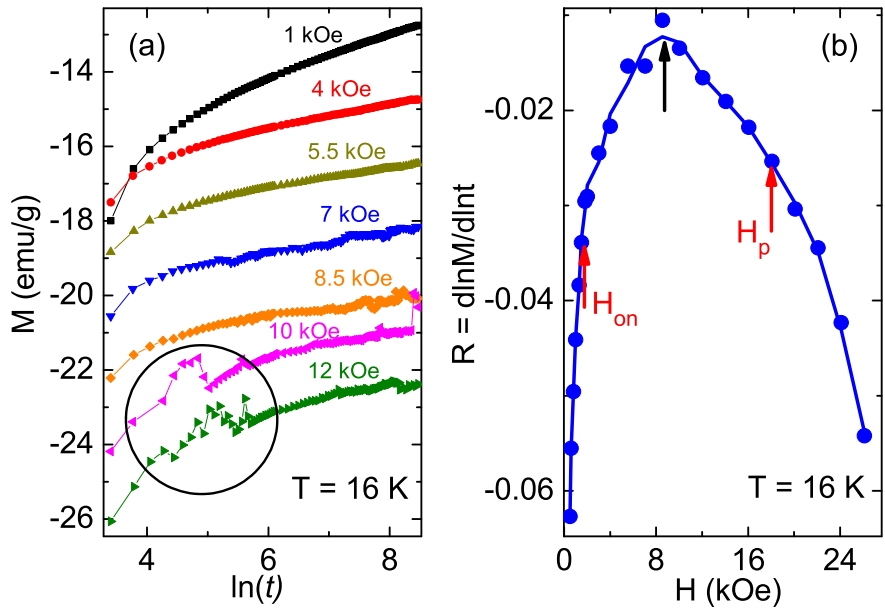}
\caption{\label{fig:Mt_RH} (a) Magnetic relaxation at T = 16 K measured under different magnetic fields. The circle shows the noise observed for magnetic fields near the peak in $R(H)$ for the same temperature. (b) Magnetic field dependence of the relaxation rate for T = 16K.}
\end{figure}

\begin{figure}[h]
\centering
\includegraphics[height=8cm]{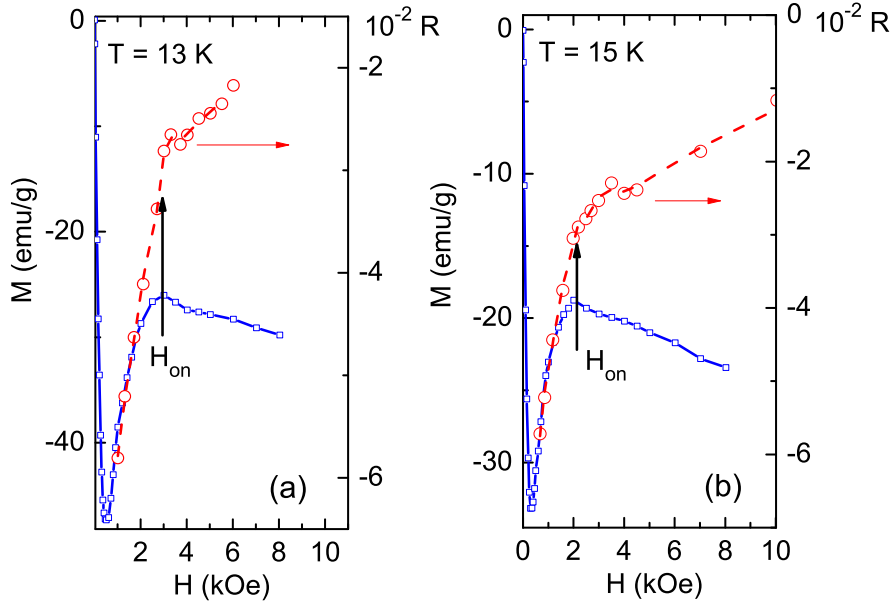}
\caption{\label{fig:MH_RH} Isothermal magnetic field dependence of magnetization and relaxation rate for $T$ = 13 K and 15 K. In both panels, vertical arrows show a sharp peak in H$_{on}$ and the associated change of slope in the magnetic relaxation rate ($R$).}
\end{figure}

Figure \ref{fig:Mt_RH}a shows the isothermal magnetic relaxation data at $T$ = 16 K measured for different magnetic fields, and Fig.\ref{fig:Mt_RH}b shows $R(H)$ for $T$ = 16 K. It is interesting to observe that the magnetic relaxation data is quite smooth for low magnetic fields but show a noisy behavior as the field approaches the peak in $R(H)$ for the same temperature. A similar behaviour has also been observed in the magnetic relaxation data measured at other temperatures as well (not shown). Since the peak in $R(H)$ (between H$_{on}$ and H$_p$) suggests a vortex lattice structural phase-transition, the noise in the experimental data (Fig. \ref{fig:Mt_RH}) may have the same origin. However, such explanation must be taken with caution, since the vortex lattice does not show long range order in such compounds \cite{esk09, yin09, vin09}. Another interesting observation (shown in Fig. \ref{fig:MH_RH}) is that the onset of the SMP (at H$_{on}$) in measured isothermal $M(H)$ is quite sharp, which has not been observed before. It has also been noticed that $R(H)$ changes its slope in the vicinity of H$_{on}$. This may be explained in terms of the pinning crossover which renders a peak in $M(H)$, as has been also discussed in our recent study \cite{shy17}.   

\begin{figure}[h]
\centering
\includegraphics[height=8cm]{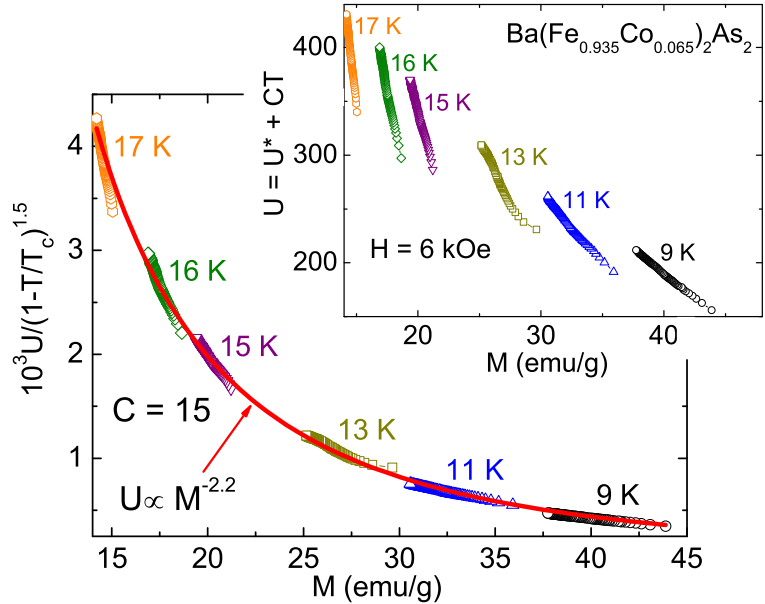}
\caption{\label{fig:Maley} Activation energy $U(M)$ after scaling with the function $g(T/T_c)=(1-T/T_c)^{1.5}$ for $H$ = 6 kOe. Inset: $U(M)$ before scaling.}
\end{figure}

To study such a possible pinning crossover, we obtained the activation energy ($U$) as a function of magnetic moment ($M$) from the $M(t)$ data, by using the approach developed by Maley's et. al. \cite{mal90} and later exploited in many recent studies \cite{shy17, wei16, said10}.

\begin{equation} \label{eq:1}
U = -Tln[dM(t)/dt] + CT,
\end{equation}

where C is a constant depending on the hoping distance of the vortex, the attempt frequency and the sample size. The inset of Fig. \ref{fig:Maley}, shows  $U(M)$ for $H$ = 6 kOe in the temperature range from 9 K to 17 K (in the SMP region, see the $H$-$T$ phase diagram). For each temperature, the activation energy was obtained by using C = 15, which is commensurate with the values presented in the literature \cite{hen91}. As it is observed that the inset of Fig. \ref{fig:Maley} do not shows the smooth behaviour of $U$ vs. $M$ curve. Hence, to obtain the smooth curve of $U(M)$, we scaled $U$ with the function $g(T/T_c) = (1-T/T_c)^{1.5}$, as suggested by McHenry et. al. \cite{hen91}. The main panel of Fig. \ref{fig:Maley}, shows the smooth curve of $U(M)$ after scaling and  follow the power law behaviour. In this analysis, best smooth curve of $U(M)$ was used for C = 15, which is further exploited to estimate the activation energy from the different $M(t)$ curves measured in different magnetic field regimes ($H<H_{on}$, $H_{on}<H<H_p$ and $H>H_p$) at $T$ = 15 K.

\begin{figure}[h]
\centering
\includegraphics[height=8cm]{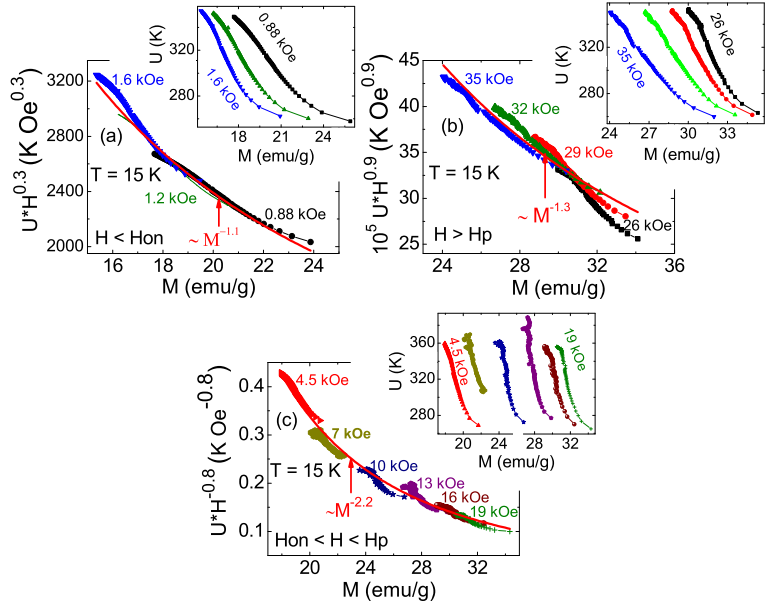}
\caption{\label{fig:CPT} Insets of each panel shows the $U(M)$ obtained through $M(t)$ measured at $T$ = 15 K, in presence of different magnetic fields. The main panels show the $U(M)$ curve scaled according to the collective creep theory.}
\end{figure}

To study the creep mechanism above, below, and in the intermediate field regime of SMP, we plotted the $U(M)$ curves in the inset of each panel of Fig. \ref {fig:CPT} for magnetic fields $H<H_{on}$, $H>H_p$ and $H_{on}<H<H_p$ at $T$ = 15 K. The activation energy shown in each inset of Fig. \ref {fig:CPT} is analysed using the collective creep theory \cite{fei89, abu96}, in which, $U(B,J) = B^{\nu}J^{-\mu}$ $\approx$ $H^{\nu}M^{-\mu}$, where the critical exponents $\nu$ and $\mu$ depend on the specific pinning regime. It has been shown that the activation energy increases (decreases) with magnetic field for collective (plastic) creep \cite{abu96}. Hence, the expression for $U(B,J)$, mentioned above, suggests a positive (negative) value of exponent $\nu$ for collective (plastic) creep. Each panel of Fig. \ref {fig:CPT} shows the smooth behaviour of $U(M)$ after scaling with different values of $\nu$. It is interesting to observe that each smooth curve of $U(M)$ follows a power law. The scaling of the $U(M)$ curves with $H$, clearly suggests the collective creep nature in the regime $H_{on}<H<H_p$ (Fig \ref {fig:CPT}c), which changes to plastic creep above $H_p$ (Fig \ref {fig:CPT}b). This result unambiguously demonstrate that the SMP in the present case is due to a crossover in the creep mechanism from collective to plastic. The scaling of $U(M)$ in Fig. \ref {fig:CPT}a, suggests the plastic nature of creep for $H<H_{on}$. However, similar behaviour of scaling below $H_{on}$ has also been observed in many studies which explained this behaviour in terms of single vortex pinning (SVP) \cite{shy17}. The change of SVP to collective creep above $H>H_{on}$ also shows a peak at $H_{on}$ in $M(H)$ which is quite different than the SMP observed at $H_p$. 

A similar, elastic (collective) to plastic creep crossover has been also reported to explain the SMP in other pnictide superconductors of the 122 family, such as K-doped BaFe$_2$As$_2$ \cite{shy17, said10}, Na-doped BaFe$_2$As$_2$ \cite{pra13}. However, such pinning crossover is also observed through magnetic relaxation measurements in  Ca$_{0.25}$Na$_{0.75}$Fe$_2$As$_2$ 122 superconductor \cite{hab11}, but no SMP exists in the sample. Recently, such a crossover from elastic to plastic is also claimed in a new superconductor (Li$_{1−x}$Fe$_x$)OHFeSe, that do not show the SMP \cite{chu17}. Also, some  superconductors of the 122 family, such as Ca(Fe$_{1-x}$Co$_x$)$_2$As$_2$ (x= 0.056) does not show the SMP \cite{pra10}and its vortex dynamics is described in terms of plastic creep theory. The reason for SMP in Ni-doped BaFe$_2$As$_2$ 122 superconductor is still under investigation \cite{said13, said11}. 

\begin{figure}[h]
\centering
\includegraphics[height=8cm]{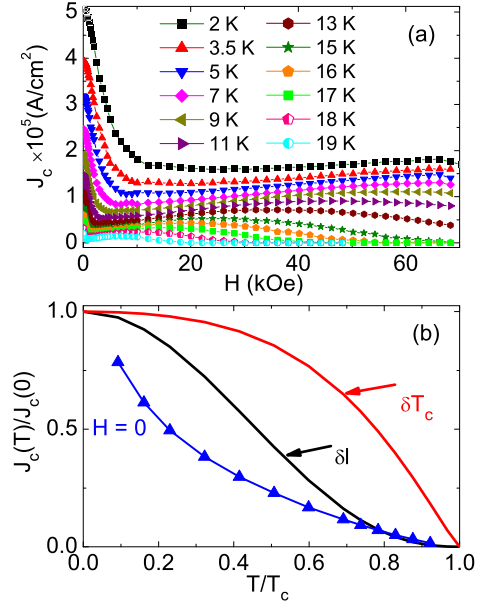}
\caption{\label{fig:Jc} (a) $H$ dependence of $J_c$ for different temperatures, as follows from Bean's critical state model. (b) Reduced temperature ($T/T_c$) dependence of $J_c(T)/J_c(0)$ in zero applied magnetic field. The lines correspond to the indicated pinning models.}
\end{figure}

Figure \ref{fig:Jc}a shows the $H$ dependence of the critical current density ($J_c(H)$) at different temperatures as estimated using the Bean's critical state model \cite{bean64} through $J_c (A/cm^2) = 20 \Delta M/a(1-a/3b)$, where $\Delta M$ (emu/cm$^3$) is the difference in the field-increasing and field-decreasing branches of isothermal $M(H)$, and $a$, $b$ are the dimensions (cm) of the sample perpendicular to the magnetic field. The $J_c$ value at $T$ = 2 K in the zero magnetic field limit is about 5 $\times$ 10$^5$ A/cm$^2$. A similar value of $J_c$ has also been observed for the Ba(Fe$_{0.9}$Co$_{0.1}$)$_2$As$_2$ and Ba(Fe$_{0.93}$Co$_{0.07}$)$_2$As$_2$ single crystals at $T$ = 5 K in the zero magnetic field limit \cite{pro08, yas09}. This $J_c$ value is of same order of magnitude as for K-doped BaFe$_2$As$_2$ superconductors \cite{shy17, son16} and may be exploited for future technological purpose. However, in thin films of Co-doped BaFe$_2$As$_2$ superconductors, the observed $J_c$ value is more than 10$^6$ A/cm$^2$ at $T$ = 4.2 K \cite{tak11b, tak11a, tak10}. Recently, high $J_c$ was reported in P-doped BaFe$_2$As$_2$ thin film \cite{kur15}. In the literature, pnictide superconductors of the 122 family are considered as the most promising for practical usage \cite{lla15}. Fig. \ref{fig:Jc}b shows $J_c(T)/J_c(0)$ in zero magnetic field as a function of the reduced temperature ($T/T_c$) to explore the pinning mechanism using the model proposed by Griessen et. al \cite{gri94}. It clearly shows that at low temperatures the pinning does not follow the $\delta l$ ($J_c(t)/J_c(0)=(1-t^2)^{5/2}(1+t^2)^{-1/2}$) and $\delta T_c$ ($J_c(t)/J_c(0)=(1-t^2)^{7/6}(1+t)^{5/6}$) models. A similar pinning behaviour has already been observed in our previous study on Ba$_{0.75}$K$_{0.25}$Fe$_2$As$_2$ single crystal \cite{shy17} and in many other pnictide superconductors of 122 family, which is argued in terms of unconventional intrinsic pinning \cite{vla15}. However, above 15 K (t = 0.69), the $J_c(T)/J_c(0)$ curve shows its resemblance with the $\delta$l pinning behaviour, which suggests that at high temperature (below $T_c$) the pinning is related to the spatial variations of the charge carrier mean free path. A detailed description about the $\delta$l and $\delta T_c$ pinning in high $T_c$ superconductors is provided in Ref. \cite{kob00}.

\begin{figure}[h]
\centering
\includegraphics[height=8cm]{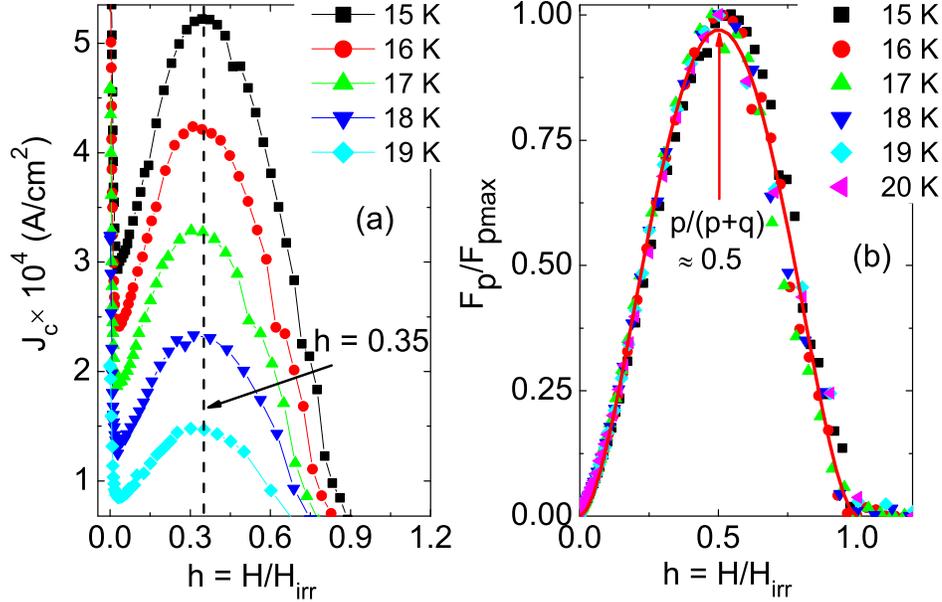}
\caption{\label{fig:Fp} (a) Critical current density ($J_c$) as a function of the reduced field ($h$) at different temperatures (15 K $\leq T<T_c$). The maximum in each curve is shown at $h$= 0.35. (b) Normalized pinning force density ($F_p/F_{p-max}$) as a function of reduced magnetic field ($h$) at different constant temperatures. The data for all temperatures collapse in one single curve and showing a maximum at $h$ = 0.5. The solid red line is the best fit of the Dew-Hughes model (see main text for details).}
\end{figure}

To further explore the pinning mechanism in the sample, $J_c$ and the normalized pinning force density $F_p/F_{p-max}$ are plotted as a function of the reduced magnetic field ($h$=$H/H_{irr}$) in Fig. \ref{fig:Fp} a,b respectively. It is commonly considered and has also been seen by Fang et. al. \cite{fan11} that the maximum in $F_p$ is responsible for the maximum in $J_c$ (or SMP). However, in the present case, it is interesting that the maximum in $J_c$ (Fig. \ref{fig:Fp}a) lies at about $h$ = 0.35, whereas, the maximum in $F_p$ (Fig. \ref{fig:Fp}b) is found at $h$ = 0.5. It suggests that the maximum in $J_c$ or the SMP in the sample is not directly related with the maximum in the pinning force density. This important issue needs to be sort out in future investigations. The normalized pinning force curves ($F_p/F_{p-max}$) vs. $h$ shown in Fig. \ref{fig:Fp} b has been analysed using the pinning model suggested by Dew-Hughes \cite{dew74}, in which, the normalized pinning force density may follow a general mathematical form, $F_p/F_{p-max} = A h^p (1-h)^q$, where, $A$ is multiplicative factor, $F_{p-max}$, is the maximum pinning force density at constant temperature and the parameters $p$ and $q$ provide the details about the pinning mechanism \cite{mic16}. The above expression leads to a single peak behaviour in the case of single dominating pinning mechanism. In the present case, the functional form of the fitted function is found as $F_p/F_{p-max} = 12.46 h^{1.85} (1-h)^{1.84}$, which is shown in Fig. \ref{fig:Fp}b. It should be noted that the data show a good scaling with $h$ and a single peak behaviour at $h$ = 0.5, which suggests a single dominating pinning mechanism in the sample. As suggested by Dew-Hughes \cite{dew74} the peak at $h$ = 0.5 of the scaled pinning force curves suggests a $\Delta \kappa$-type pinning, however, for such pinning to exist, $p$ and $q$ parameters should be equal to 1.

\section{Conclusion}
A detailed study of isothermal magnetic field dependence of the magnetization, $M(H)$, and the magnetic relaxation, $M(t)$, on a Ba(Fe$_{0.935}$Co$_{0.065}$)$_2$As$_2$ single crystal is presented. Below $T_c$ (= 21.7 K), the isothermal $M(H)$ shows the clear signature of the second magnetization peak (SMP) with a sharp H$_{on}$. The field dependence of the relaxation rate $R$ measured for different isothermals shows a peak (H$_{spt}$) between H$_{on}$ and H$_p$. In the literature, the peak in $R(H)$ is related to a rhombic to sqaure vortex-lattice structural phase transition. The magnetic relaxation data measured for magnetic fields near H$_{spt}$ shows a noisy behavior, which might be related to such phase transition. However, literature suggests that the vortex lattice in Co-doped BaFe$_2$As$_2$ superconductors is highly disordered in nature and masks the observation of the structural phase transition. More direct evidences are required in future studies to observe such phase transition in Ba(Fe, Co)$_2$As$_2$ superconductors. The magnetic relaxation data is used to obtain the activation energy ($U$) and is analysed using the Maley's method and collective pinning theory. The analysis convincingly shows that the SMP is due to the collective (elastic) to plastic creep crossover which accompanies the structural phase transition occuring below $H_p$. The pinning properties in the sample are explored using the models developed by Dew-Hughes and Griessen et. al. It suggests a single dominating pinning mechanism which is different than the conventional $\delta$l and $\delta T_c$ pinning behaviours. The critical current density ($J_c$) is estimated using the Bean's critical state model and found to be about 5 $\times$ 10$^5$ A/cm$^2$ at $T$ = 2 K in the zero magnetic field limit, which is comparable to other Co-doped pnictide superconductors. Interestingly, the maximum critical current density in the sample is not directly related to the maximum in the pinning force density ($F_p$). 

\section*{Acknowledgements}

SS acknowledges a fellowship from FAPERJ (Rio de Janeiro, Brazil), processo: E-26/202.848/2016, and also would like to thank Prof. Luis Ghivelder for his constant support \& encouragement during the work. JM and DS acknowledge financial support from project FIS2016-79109-P (AEI/FEDER, UE), and from the Xunta de Galicia (project AGRUP 2015/11). SSS and ADA acknowledges support from CNPq. The work at Oak Ridge National Laboratory was supported by the U.S. Department of Energy (DOE), Office of Science, Basic Energy Sciences (BES), Materials Science and Engineering Division.

\section*{References} 

\bibliographystyle{iopart-num}


\providecommand{\newblock}{}

\end{document}